\documentclass[intlimits,twoside,a4paper]{article}

\usepackage{amsmath,amssymb}
\usepackage{graphicx}
\usepackage{wrapfig}

\usepackage[T2A]{fontenc}
\usepackage[cp1251]{inputenc}

\usepackage[eqsecnum]{cmpj2}

\issue{2012}{15}{4}{43601}

\doinumber{10.5488/CMP.15.43601}

\title[On the effect of superfluid flows on the interaction of microwaves with He II]%
{On the effect of superfluid flows on the interaction \\ of microwaves with He II}

\author[V.D. Khodusov, A.S. Naumovets]{V.D. Khodusov, A.S. Naumovets}
\address{Karazin Kharkov National University, 4 Svobody Sq., 61077 Kharkov, Ukraine}

\authorcopyright{V.D. Khodusov, A.S. Naumovets, 2012}

\date{Received July 3, 2012, in final form August 22, 2012}

\begin{document}

\maketitle

\begin{abstract}
The paper proposes a possible mechanism of interaction of microwaves with superfluid helium that results in an experimentally observed narrow peak of microwave absorption on the frequencies by the order of the roton frequency. The obtained microwave photon absorption coefficient depends on the local equilibrium distribution function which is established due to fast roton-roton and roton-phonon interactions. With the availability of superfluid flows, the local equilibrium distribution function depends on their velocity. The critical velocity of the flows, at which the absorption of microwaves is replaced by their radiation, is found.
\pacs 67.25.du, 67.30.eh, 67.25.dt
\end{abstract}

\section{Introduction}

The experiments aimed at the study of the absorption of microwaves in the frequency range $40\div200$~GHz in superfluid helium have produced a number of unexpected results. Thus, a resonant absorption of microwaves has been revealed in the frequencies relevant to roton energy~\cite{3,4,5,6}; in this case, a narrow absorption line near the minimum energy of rotons has been observed against the background of a wide pedestal and, what is more, the results of the measurements of temperature dependence of this narrow line correlated with the change in the minimum roton energy. Besides, the effect of the velocity of the relative motion of normal ${\bf V}_\mathrm{n}$  and superfluid ${\bf V}_\mathrm{s}$ components on the character of the resonant interaction of microwaves has been found. The microwave absorption coefficient decreased at an increase of the velocity of the relative motion. When there was an excess of some critical velocity, the absorption of waves was replaced by their radiation.Similar effects take place in plasma physics when one describes linear and nonlinear Landau attenuation of plasma waves on particles as well as the excitation of these waves by particle streams moving at the velocity higher than some critical value~\cite{7}.

These experimental results are indicative of an intensified manifestation of electric properties in helium at temperatures below \mbox{$T_\lambda$}. Moreover, they point out a special role, which the quasi-particles (i.e., phonons, rotons) should play in explaining these experiments. In~\cite{9,9a,10,11,12,13,14}, some attempts were made to theoretically explain these features. In the above works, a possible mechanism of microwave interaction with superfluid helium is proposed. It consists in taking into account the effect of its electric properties on quasi-particles and their kinetics. In the temperature range, in which the experiments were carried out ($1.4 \div 2.3$~K), rotons had a dominant role. They determined both thermodynamic and kinetic properties of He II. Fast roton-roton and roton-phonon interactions provide the establishment of a hydrodynamic regime in gas of quasi-particles. The explanation of these effects can be made if we take into account the summands in the roton energy that are linear in the electric field~\cite{15}, i.e., if one suggests that a roton has a dynamic dipole moment.

\section{Raman scattering of electromagnetic waves from quasiparticles in He~II}

The interaction of electromagnetic waves with superfluid helium had been previously studied, both theoretically and experimentally, while describing the Raman light scattering from phonons, second sound and rotons~\cite{16,17,18,19,20,21}. In references~\cite{17,18}, the two-roton scattering of light was experimentally observed. Since the roton momentum considerably exceeds (by some orders) the photon momentum (visible light and microwaves)~\cite{19}, in order to monitor the electromagnetic wave Raman scattering from rotons, it is necessary to have two rotons, which follows from the momentum conservation. Energy and momentum conservation permits a birth of two rotons by photon and photon scattering from rotons.

For the first process, the energy and momentum conservation laws are:
\begin{equation} \label{1conslow}
\hbar \omega_1=\hbar \omega_2+\varepsilon_3+\varepsilon_4 \, , \qquad \hbar {\bf k}_1=\hbar {\bf k}_2+{\bf p}_3+{\bf p}_4 \, ,
\end{equation}
where $\omega_{1,2}$ and ${\bf k}_{1,2}$ are the frequencies and wave vectors of the incident and the reflected electromagnetic wave accordingly, $\varepsilon_{3,4}$  and ${\bf p}_{3,4}$ are the energies and the momenta of rotons, respectively. The scattering in this case occurs with the excitation of two rotons with the opposite momenta  ${\bf p}_3\approx-{\bf p}_4$.

Taking into account that the roton energy is $\varepsilon=\Delta+{(p-p_0)^2}/{2\mu}$ , from \eqref{1conslow} one can  obtain for Stokes (red) satellite:
\begin{equation} \label{redsatel}
\hbar (\omega_1-\omega_2)\geqslant 2\Delta+\frac{(p_3-p_0)^2}{\mu} \, ,
\end{equation}
where $p_0$ is the value of a momentum, at which the energy or roton has a minimum equal to $\Delta$, $\mu$ is the effective mass of roton.
The density of the number of roton energy states per a unit of volume is equal to:
\begin{equation} \label{density}
\rho(\varepsilon)=\frac{\sqrt{2\mu}}{4\pi^4\hbar^3}\frac{\left[p_0+\sqrt{2 \mu (\varepsilon-\Delta)}\right]^2}{\sqrt{\varepsilon-\Delta}} \, .
\end{equation}
From equation~\eqref{density} it follows that the basic role in these processes is played by rotons having minimal energy $\varepsilon\approx\Delta$ .
Then, as follows from equation~\eqref{redsatel}, two-roton light absorption takes place, as observed in works~\cite{17} and~\cite{18}. Theoretical explanation of these experiments as well as the detection of matrix elements of photon and roton interactions are given in references~\cite{19} and~\cite{20}.
For the second process, describing photon scattering from rotons, conservation laws of energy and momentum are:
\begin{equation} \label{2conslow}
\hbar \omega_1+\varepsilon_3=\hbar \omega_2+\varepsilon_4 \, , \qquad \hbar {\bf k}_1+{\bf p}_3=\hbar {\bf k}_2+{\bf p}_4 \, .
\end{equation}
In contrast to the first case, here we have ${\bf p}_3\approx{\bf p}_4$.

As follows from the compatibility condition of the system \eqref{2conslow}, the next condition is imposed on frequencies of the incident and the reflected electromagnetic waves, for Stokes satellite:
\begin{equation}\label{Stokessatel}
0\leqslant \left(\omega _{1} -\omega _{2} \right)\leqslant \frac{2\left({ p}_{3} -{ p}_{0} \right)\left|{\bf k}_{1} -{\bf k}_{2} \right|+\hbar \left|{\bf k}_{1} -{\bf k}_{2} \right|^{2} }{2\mu } \, .
\end{equation}

This condition, as opposed to \eqref{redsatel}, determines an upper limit for $\left(\omega _{1} -\omega _{2} \right)$  and shows that frequencies $\omega _{1}$  and  $\omega _{2}$ are closer to each other than in the previous case. Indeed, under the conditions corresponding to those for the experiments in~\cite{17},  estimations give the following result for $p_{3} \approx p_{0} $, ${\bf k}_{1} \approx -{\bf k}_{2} $, $k_{1} \approx 10^{5}$~cm$^{-1}$: $\left(\omega _{1} -\omega _{2} \right)\approx 2\cdot 10^{6}$~s$^{-1}$. In the first case, under the same conditions, from equation~\eqref{redsatel} we obtain $\left(\omega _{1} -\omega _{2} \right)\approx 2.27\cdot 10^{12}$~s$^{-1}$, if we take the value for the roton minimal energy at temperature $T=1.4$~K $\Delta =8.65$~K. Stokes line intensity of Raman light scattering on rotons due to the second process, as it follows from the research~\cite{20}, is $\re^{-{\Delta }/{T} } $  times less than that one due the first process. Moreover, it is in other frequency ranges (MHz). We hope that with the use of modern equipment it will be possible to detect this line in spite of small intensity.
A method of resonant combinational light scattering represents a special interest in studying Raman scattering~\cite{22}. Thus, the instances are possible when light frequency coincides with the own frequencies of elementary excitations. In this case, there is an imposition of two effects: a forced resonant excitation of quasi-particles by electromagnetic wave and photon scattering from quasi-particles. If an electromagnetic wave frequency tends to roton energy, the processes of the birth of two rotons are forbidden, as it follows from energy conservation.

\section{The narrow resonant line and the effect of the flows}

We assume that a narrow resonant line, which is observed on the background of the pedestal (Rayleigh wings), is caused by the photons scattering from rotons. The quasi-local distribution function of rotons in this case is established in a time $1/\gamma _{r} \sim 10^{-11}$~s  as:
\begin{equation}\label{Nfunc}
n_{r0} =\left[\exp \left(\frac{\varepsilon +{\bf p}{\bf W}}{T} \right)-1\right]^{-1}\, ,
\end{equation}
where ${\bf W}={\bf V}_\mathrm{n} -{\bf V}_\mathrm{s} $ is the relative velocity.

The change in a unit of time of a number of microwave photons $\Delta N_{1}$ with energies
$\hbar \omega _{1}$ due to the induced processes of scattering of photons from rotons can be written in the form:
\begin{eqnarray}\label{Nfuncspeed}
 \frac{\partial \Delta N_{1} }{\partial t} &=&\Delta N_{1} \int \left|\Phi \left(1,3;2,4\right)\right| ^{2} N_{2} \left(n_{04} -n_{03} \right)\delta \left(\hbar \left(\omega _{1} -\omega _{2} \right)+\varepsilon _{3} -\varepsilon _{4} \right)  \\
 &\times& \delta \left({\bf p}_{4} -{\bf p}_{3} -\hbar \left({\bf k}_{1} -{\bf k}_{2} \right)\right)\frac{\rd^{3} p_{4} \rd^{3} p_{3} }{\left(2\pi \hbar \right)^{6} } \frac{\rd^{3} k_{2} }{\left(2\pi \right)^{3} }\, .\nonumber
\end{eqnarray}
$\left|\Phi \left(1,3;2,4\right)\right|$ is a matrix element of interaction of rotons and microwaves. It depends from an interaction constant, which determines the interaction between the dipole moment of rotons and the electric component of microwaves. This constant is small, because  the electromagnetic field relaxation occurs during the period of an order of seconds after  turning off the field~\cite{privat}.
Taking into account that the momentum of photons is much smaller than that of the rotons and ${\bf p}_3\approx{\bf p}_4$, we can write $n_{04}$ as:
\begin{equation} \label{nexpand}
n_{04} =n_{03}+\frac{n_{03} }{T} \left[-\hbar \left(\omega _{1} -\omega _{2} \right)+\hbar \left({\bf k}_{1} -{\bf k}_{2} \right){\bf W}\right].
\end{equation}

If we try to find a solution to this equation in the form $\Delta N\sim \re^{-\gamma t} $, we obtain
\begin{eqnarray} \label{decr}
\gamma &=&-\frac{1}{T} \int \left|\Phi \left(1,3;2,4\right)\right| ^{2} n_{03} N_{2} \left[(\omega _{1} -\omega _{2} )-\left(k_{1} -k_{2} \right)W\right] \\
&\times &\frac{2\mu }{\left|k_{1} -k_{2} \right|\sqrt{(p_{3} -p_{0} )^{2} +2\mu \hbar (\omega _{1} -\omega _{2} )} } \frac{2\pi p_{3}^{2} \rd p_{3} }{\left(2\pi \hbar \right)^{3} } \frac{\rd^{3} k_{2} }{\left(2\pi \right)^{3} }\, . \nonumber
\end{eqnarray}
 By comparing this attenuation factor with the one experimentally measured, it is possible to obtain the matrix element order estimates of interaction of microwave photons and rotons.
The absorption coefficient of microwaves depends upon $\omega _{1}$. Assuming that the frequencies  $\omega _{1}$, $\omega _{2}$ fall into the same range in the vicinity of the resonance, it is possible to determine the boundary conditions from \eqref{Stokessatel}:
\[\frac{\Delta }{\hbar } -\frac{2\left|p_{3} -p_{0} \right|\left|k_{1} -k_{2} \right|+\hbar \left|k_{1} -k_{2} \right|^{2} }{4\mu } \leqslant \omega _{1} \leqslant \frac{\Delta }{\hbar } +\frac{2\left|p_{3} -p_{0} \right|\left|k_{1} -k_{2} \right|+\hbar \left|k_{1} -k_{2} \right|^{2} }{4\mu }\, . \]

From this relation it follows, that both the resonant frequency itself and the limiting values of the permitted frequencies of microwaves change as a function of temperature basically in the same way as the energy gap in a roton spectrum, which corresponds to the experimental results.

Besides, if ${\bf k}_{1} \approx -{\bf k}_{2} $ and $\left|p_{3} -p_{0} \right|\approx \sqrt{2\mu T} $  for the thermal rotons, the width of a resonant curve at zero intensity is defined by the expression:
\begin{equation} \label{width}
\Delta \omega \approx 2{\left(\sqrt{2\mu T} k_{1} +\hbar k_{1} ^{2} \right)\Big/ \mu }.
\end{equation}

If $T=1.4{\rm \; }$~K, $k_{1} \approx 37.7{\rm \; }$~cm$^{-1}$, then $\Delta \omega \approx 4.7\cdot 10^{5}$~Hz. It coincides by the order of the magnitude with the one which has been observed in the experiment.

From the expression~\eqref{decr}, for $\gamma $  it is obvious that there exists such a relative critical velocity $W_\mathrm{cr} $ at which $\gamma =0$. At the velocities greater than $W_\mathrm{cr} $, absorption of microwaves is replaced by their radiation. After the calculation of the remaining integral, we obtain the critical velocity: $W_\mathrm{cr} =\frac{8}{3} \frac{\hbar k_{1} }{\mu } $. Substituting the values of  $k_{1} $ and $\mu $, we obtain the following value of $W_\mathrm{cr} \sim 10^{-2}$~cm/s.

\section{Conclusions}

A possible mechanism of interaction between microwaves and superfluid He is proposed in the paper. The obtained microwave phonon absorption coefficient depends on the local equilibrium distribution function, which is established due to fast roton-roton and roton-phonon interactions. It is shown that the resonant line changes with the temperature similarly to the minimum roton energy. The value of the resonant line width is obtained, which is in agreement with the experimentally observed one. The local equilibrium distribution function and, correspondingly, the microwave absorption coefficient depend on the velocity of superfluid flows. There is found a critical velocity of flows at which the absorption of microwaves is replaced by their radiation.


\vspace{-5mm}

\ukrainianpart

\title{Вплив надплинних потоків на взаємодію надвисокочастотних хвиль з Не ІІ}
\author{В.Д. Ходусов, А.С. Наумовець}
\address{Харківський національний університет ім. В.Н. Каразіна, пл. Свободи, 4, 61077 Харків, Україна
}

\makeukrtitle

\begin{abstract}
\tolerance=3000%
В роботі запропоновано механізм взаємодії надвисокочастотних хвиль із надплинним гелієм, який пояснює експериментально спостережуваний вузький пік поглинання надвисокочастотних хвиль з частотою порядку ротонної частоти. Отриманий коефіцієнт поглинання надвисокочастотних фотонів залежить від локально-рівноважної функції розподілу, яка встановлюється завдяки швидких ротон-ротонних та ротон-фононних взаємодій. При наявності надплинних потоків, локально-рівноважна функція розподілу залежить від їхньої швидкості. Отримано критичну швидкість потоків, за якої поглинання надвисокочастотних хвиль змінюється на їхнє випромінювання.
\keywords
надплинність, ротони, надвисокочастотні хвилі

\end{abstract}

\lastpage

\end{document}